\documentclass[draftcls,onecolumn]{IEEEtran}

\usepackage{pifont,calc}
\usepackage{cite}

\usepackage{graphicx}
\ifCLASSINFOpdf
   \usepackage[pdftex]{graphicx}
\else
 \fi
\usepackage[cmex10]{amsmath}
\usepackage{array}

\usepackage{cases} 
\usepackage{bm} 

\makeatletter

\newcommand{\Rmnum}[1]{\expandafter\@slowromancap\romannumeral #1@}
\makeatother

\begin{document}

\title{An Improved Square-root  Algorithm for V-BLAST Based on Efficient Inverse Cholesky Factorization}

\author{Hufei Zhu,~
        Wen Chen,~
        Bin Li,~
        Feifei Gao,~\IEEEmembership{Members,~IEEE}
\thanks{H. Zhu and B. Li are with Huawei Technologies Co., Ltd.,
Shenzhen 518129, P. R. China, e-mail: \{zhuhufei;binli\}@huawei.com.}
\thanks{W. Chen is with Dept. of Electronic Engineering, Shanghai Jiao Tong Univ., Shanghai 200240,
P. R. China, e-mail: wenchen@sjtu.edu.cn.}
\thanks{F. Gao is with the School of Engineering and Science, Jacobs University, Bremen, Germany, 28759.
Email: feifeigao@ieee.org. }
\thanks{This work is supported by NSF China \#60972031, by SEU SKL project
\#W200907, by Huawei Funding \#YJCB2009024WL and \#YJCB2008048WL,
and by National 973 project \#2009CB824900, by National Huge Project
\# 2009ZX03003-002 of China.}
}

\markboth{IEEE Transactions on Wireless Communications}%
{Shell \MakeLowercase{\textit{et al.}}: Bare Demo of IEEEtran.cls for Journals}

\maketitle

\begin{abstract}
A fast algorithm for inverse Cholesky factorization is proposed, to
compute a triangular square-root of the estimation error covariance
matrix for Vertical Bell Laboratories Layered Space-Time
architecture (V-BLAST). It is then applied to propose an improved
square-root algorithm for V-BLAST, which speedups several steps in
the previous one, and can offer further computational savings in
MIMO Orthogonal Frequency Division Multiplexing (OFDM) systems.
Compared to the conventional inverse Cholesky factorization, the
proposed one avoids the back substitution (of the Cholesky factor),
and then requires only half divisions. The proposed V-BLAST
algorithm is faster than the existing efficient V-BLAST algorithms.
The expected speedups of the proposed square-root V-BLAST algorithm
over the previous one and the fastest known recursive V-BLAST
algorithm are $3.9\sim5.2$ and $1.05\sim1.4$, respectively.
\end{abstract}


\IEEEpeerreviewmaketitle

\section{Introduction}

\IEEEPARstart{M}{ultiple}-input multiple-output (MIMO) wireless
communication systems can achieve huge channel capacities
\cite{Jun28_MIMO} in rich multi-path environments through exploiting
the extra spatial dimension. Bell Labs Layered Space-Time architecture (BLAST) \cite{BLASTfeb25},
including the relative simple
vertical BLAST (V-BLAST) \cite{zhf1},
is such a system that maximizes the data rate by transmitting
independent data streams simultaneously from multiple antennas.
V-BLAST often adopts the ordered successive interference
cancellation (OSIC) detector~\cite{zhf1},
which detects the data streams
iteratively with the optimal ordering. In each iteration,
the data stream with the highest signal-to-noise ratio (SNR) among
all undetected data streams is detected through a zero-forcing
(ZF) or minimum mean-square error (MMSE) filter. Then the effect
of the detected data stream is subtracted from the received signal vector. 

Some fast algorithms have been proposed
\cite{zhf2,vtc08_zhfVTC4,vtc08_reviewer_a,zhf3,zhf5,zhf4,reviewer_VTC08_b,zhf6}
to reduce the computational complexity of the OSIC V-BLAST
detector~\cite{zhf1}.
An efficient square-root
algorithm was proposed in \cite{zhf2} and then improved in
\cite{vtc08_zhfVTC4}, which also partially inspired the modified
decorrelating decision-feedback algorithm \cite{vtc08_reviewer_a}.
In additon, a fast recursive algorithm was proposed in \cite{zhf3} and then improved in \cite{reviewer_VTC08_b,zhf6,zhf5,zhf4}. 
The improved recursive algorithm 
in
\cite{reviewer_VTC08_b}
requires less multiplications and more
additions than the original recursive algorithm \cite{zhf3}. In \cite{zhf6}, the author gave the
``fastest known algorithm" by incorporating improvements proposed in
\cite{zhf5,zhf4} for different parts of the original recursive algorithm
\cite{zhf3}, and then proposed a further improvement for the
``fastest known algorithm".

On the other hand, most future cellular wireless standards are based on 
MIMO Orthogonal
Frequency Division Multiplexing (OFDM) systems,
where the OSIC V-BLAST detectors
\cite{zhf1,zhf2,vtc08_zhfVTC4,vtc08_reviewer_a,zhf3,zhf5,zhf4,reviewer_VTC08_b,zhf6}
require an excessive
complexity 
to update the detection
ordering and the nulling vectors for each 
subcarrier. Then simplified V-BLAST detectors with some performance degradation
are proposed in
\cite{OFDM2_zhf_VTC08_8,zhf_VTC08_8}, which update the detection \cite{OFDM2_zhf_VTC08_8} or the detection
ordering  \cite{zhf_VTC08_8} per group of subcarriers to reduce the required complexity.

In this letter, a  fast algorithm for inverse Cholesky
factorization~\cite{zhf_VTC08_6} is deduced
to
compute a triangular
square-root of the estimation error covariance matrix for V-BLAST.
Then it is employed to propose an improved square-root V-BLAST
algorithm,
 which speedups several steps in the previous
square-root V-BLAST algorithm \cite{vtc08_zhfVTC4}, and 
can offer further
computational savings
in MIMO OFDM systems.


This letter is organized as follows. 
Section \Rmnum{2} describes the
V-BLAST system model.  
Section \Rmnum{3} introduces the previous
square-root algorithm \cite{vtc08_zhfVTC4} for V-BLAST. In Section \Rmnum{4}, we deduce a
 fast algorithm for inverse Cholesky factorization. 
Then in Section \Rmnum{5}, we employ it to propose an improved
square-root  algorithm for V-BLAST. Section \Rmnum{6} evaluates the
complexities of the presented 
V-BLAST algorithms. Finally, we make conclusion in
Section \Rmnum{7}.

In the following sections, $( \bullet )^T$, $ ( \bullet )^* $ and $
( \bullet )^H $ denote matrix transposition, matrix conjugate, and
matrix conjugate transposition, respectively. $ {\bf{0}}_M^{} $ is
the $M\times1$ zero column vector, while $ {\bf{{\rm I}}}_M $
 is the identity matrix of size $M$.

\section{System Model}
The considered V-BLAST system consists of $M$ transmit antennas and
$N(\ge M)$ receive antennas in a rich-scattering and flat-fading
wireless channel.
The signal vector transmitted
from $M$ antennas is ${\bf{a}}=
[a_1 ,a_2 , \cdots ,a_M ]^T$
with the covariance $ E({\bf{aa}}^H ) = \sigma _a^2
{\bf{{\rm I}}}_M $. Then the received signal vector 
\begin{equation}\label{equ:1}
{\bf{x}} = {\bf{H}} \cdot {\bf{a}} + {\bf{w}},
\end{equation}
where ${\bf{w}}$ is the $N\times 1$ complex Gaussian noise vector
with the zero mean and the covariance $\sigma _w^2 {\bf{{\rm I}}}_N$, and
\begin{displaymath}
{\bf{H}} =[{\bf{h}}_1,{\bf{h}}_2,\cdots,{\bf{h}}_M]=
[{\bf{\underline h}}_1,{\bf{\underline h}}_2,\cdots,{\bf{\underline
h}}_N ]^H
\end{displaymath}
is the $N\times M$ complex
channel matrix. 
Vectors
${\bf{h}}_m$ and ${\bf{\underline h}}_n^H$ represent the $m^{th}$
column and the $n^{th}$ row of ${\bf{H}}$, respectively.

Define $\alpha  = \sigma _w^2 /\sigma _a^2$. The linear MMSE estimate
of $\bf{a}$ is
 \begin{equation}\label{equ:2}
{\bf{\hat{a}}} = \left( {{\bf{H}}^H {\bf{H}} + \alpha {\bf{I}}_M }
\right)^{ - 1} {\bf{H}}^H {\bf{x}}.
\end{equation}
  As in \cite{zhf2,vtc08_zhfVTC4,zhf3,zhf5,zhf4,reviewer_VTC08_b,zhf6},
  we 
  focus on the MMSE OSIC detector,
  which
   usually 
  outperforms the ZF OSIC detector \cite{zhf3}. Let
\begin{equation}\label{equ:Jul3Add2}
{\bf{R}} = {\bf{H}}^H   {\bf{H}} + \alpha {\bf{I}}_M.
\end{equation}
 Then the estimation error covariance matrix \cite{zhf2}
 \begin{equation}\label{equ:3}
 {\bf{P}} = {\bf{R}}^{ - 1}=\left( {{\bf{H}}^H {\bf{H}} + \alpha {\bf{I}}_M }
\right)^{ - 1}.
\end{equation}

¡¡ The 
OSIC detection detects $M$ entries of the
transmit vector $ {\bf{a}} $
 iteratively with the optimal ordering. In each iteration, the 
entry with the highest 
SNR among all the undetected
entries is detected by a 
linear filter, and then its interference 
 is
cancelled from the received signal vector \cite{zhf1}. 
Suppose that 
the entries of 
${\bf{a}}$ are permuted
 such that the 
 detected entry is $a_{M}$, the $M^{th}$ entry.
 Then its
 interference
 is
 cancelled by
\begin{equation} \label{EqAug1IC}
 {\bf{x}}^{(M - 1)}  = {\bf{x}}^{(M)} - {\bf{h}}_M a_M,
 \end{equation}
 where $a_{M}$ is treated
 as the correctly detected entry, and the initial ${\bf{x}}^{(M)}={\bf{x}}$. Then 
 the reduced-order problem is
\begin{equation} \label{EqAug1RedOrdPrb}
{\bf{x}}^{(M - 1)} = {\bf{H}}_{M - 1} {\bf{a}}_{M - 1}  + {\bf{w}},
\end{equation} where  the deflated channel matrix ${\bf{H}}_{M - 1}  = [{\bf{h}}_1,{\bf{h}}_2\,\cdots,{\bf{h}}_{M - 1}]$, and the
reduced transmit vector ${\bf{a}}_{M - 1}  = [a_1,a_2,\cdots,a_{M - 1}]^T $. Correspondingly  we can 
deduce
the linear MMSE
estimate of ${\bf{a}}_{M - 1}$ from (\ref{EqAug1RedOrdPrb}).
The detection will proceed iteratively
until all entries are detected.

\section{The Square-Root V-BLAST Algorithms} 
The square-root V-BLAST algorithms 
\cite{zhf2},\cite{vtc08_zhfVTC4}
 calculate the MMSE nulling vectors from the matrix ${\bf{F }}$ that
  satisfies
\begin{equation}\label{equ:8}
{\bf{F }}{{\bf{F }} }^H  = {\bf{P}}.
\end{equation}
Correspondingly ${\bf{F }}$ is
a square-root matrix of  ${\bf{P}}$.
Let
\begin{equation}\label{equ:Jul3Add1}
{\bf{H}}_m=[{\bf{h}}_1,{\bf{h}}_2, \cdots,{\bf{h}}_m]
\end{equation}
denote the first $m$ columns of ${\bf{H}}$. From ${\bf{H}}_m$, we
define the corresponding ${\bf{R}}_m$, ${\bf{P}}_m$ and ${\bf{F}}_m$
by (\ref{equ:Jul3Add2}), (\ref{equ:3}) and (\ref{equ:8}),
respectively. Then the previous square-root V-BLAST algorithm in
\cite{vtc08_zhfVTC4} can be summarized as follows.

\line(1,0){240}

\qquad{\bf The Previous Square-Root V-BLAST Algorithm}

\line(1,0){240} 

{\sl Initialization}:
\begin{enumerate}
\item[{P1)}] Let $m=M$. Compute an initial $
{\bf{F }}  = {\bf{F }}_{M} $: Set ${\bf{\underline P}}_{0}^{1/2} =
({1}/{{\sqrt \alpha }}){\bf{I}}_M$. Compute ${\bf{\Pi}}_i
=\left[{\begin{array}{*{20}c}
   1 & {{\bf{\underline h}}_i^{H} {\bf{\underline P}}_{i - 1}^{1/2} }  \\
   {{\bf{0}}_M } & {{\bf{\underline P}}_{i - 1}^{1/2} }  \\
\end{array}} \right]$ and
${\bf{\Pi}}_i{\bf{\Theta }}_i  = \left[ {\begin{array}{*{20}c}
    \times  & {{\bf{0}}_M^T }  \\
    \times  & {{\bf{\underline P}}_{i}^{1/2} }  \\
\end{array}} \right]$ iteratively for $i=1,2,\cdots,N$,
where ``$\times$" denotes irrelevant entries at this time, and
${\bf{\Theta }}_i$ is any unitary transformation that block
lower-triangularizes the pre-array ${\bf{\Pi}}_i$. Finally 
${\bf{F }}={\bf{\underline P}}_{N}^{1/2}$.
\end{enumerate}

{\sl{Iterative Detection}}:
\begin{enumerate}
\item[P2)] Find the minimum length row of ${\bf{F }}_{m}$
 and permute it to the last row. Permute 
 ${\bf{a}}_{m}$
 and 
  ${\bf{H}}_{m}$
 accordingly.
\item[P3)] Block upper-triangularize ${\bf{F }}_{m}$ by
\begin{equation}\label{equ:9}
{\bf{F }}_{m} {\bf{\Sigma }} = \left[ {\begin{array}{*{20}c}
   {{\bf{F }}_{m-1} } & {{\bf{u}}_{m - 1} }  \\
   {{\bf{0}}_{m - 1}^T } & {\lambda _m }  \\
\end{array}} \right],
\end{equation} where ${\bf{\Sigma }}$ is a unitary transformation, ${\bf{u}}_{m - 1}$
 is an $(m-1)\times 1$ column vector, and $\lambda _m$
 is a scalar.
\item[P4)] Form the 
linear MMSE
estimate of $a_m$, i.e.,
\begin{equation}\label{equ:10}
{\hat{a}}_m  = \lambda _m \left[ {\begin{array}{*{20}c}
   { {\bf{u}}_{m - 1} ^H } & {\left( {\lambda _m } \right)^* }  \\
\end{array}} \right] {\bf{H}}_{m}^H {\bf{x}}^{(m)}.
\end{equation}
\item[P5)] Obtain $a_m$ from $\hat{a}_m$ via slicing.
\item[P6)] Cancel the interference 
of $a_m$ in 
${\bf{x}}^{(m)}$ by (\ref{EqAug1IC}),
to obtain the reduced-order problem (\ref{EqAug1RedOrdPrb}) with the corresponding ${\bf{x}}^{(m - 1)}$, ${\bf{a}}_{m - 1}$, ${\bf{H}}_{m - 1}$
and ${\bf{F }}_{m-1}$.
\item[P7)] If $m>1$, let $m=m-1$ and go back to step P2.
\end{enumerate}
\line(1,0){252} 

\section{A Fast Algorithm for Inverse Cholesky Factorization}
The previous square-root algorithm \cite{vtc08_zhfVTC4}
requires extremely high computational load to compute the initial
${\bf{F }}$ in step P1. So we propose a fast algorithm to compute an
initial $ {\bf{F }} $ that is upper triangular.

¡¡If $ {\bf{F }}_{m} $
 satisfies (\ref{equ:8}), any ${\bf{F }}_{m} {\bf{\Sigma }} $
 also satisfies (\ref{equ:8}). Then  there must be a square-root of ${\bf{P}}_{m}$
 in the form of  
\begin{equation}\label{equ:13}
{\bf{F }}_{m}  = \left[ {\begin{array}{*{20}c}
   {{\bf{F }}_{m-1} } & {{\bf{u}}_{m - 1} }  \\
   {{\bf{0}}_{m - 1}^T } & {\lambda _m }  \\
\end{array}} \right],
\end{equation}
as can be seen from (\ref{equ:9}). 
We apply (\ref{equ:13}) to compute $ {\bf{F }}_{m} $ from $
{\bf{F }}_{m-1} $, while the similar equation (\ref{equ:9}) is
only employed to compute $ {\bf{F }}_{m-1} $ from $ {\bf{F
}}_{m} $ in \cite{zhf2} and \cite{vtc08_zhfVTC4}.

From (\ref{equ:13}), we obtain
\begin{equation}{\label{Apd1AugEle1}}
{\bf{F}}_{m}^{ - 1}  = \left[ {\begin{array}{*{20}c}
   {{\bf{F}}_{m - 1}^{ - 1} } & { - {{{\bf{F}}_{m - 1}^{ - 1} {\bf{u}}_{m - 1} }}/{{\lambda _m }}}  \\
   {{\bf{0}}_{m - 1}^T } & {{1}/{{\lambda _m }}}  \\
\end{array}} \right].
\end{equation}
On the other hand, it can be seen that ${{\bf{R}}_{m}}$ defined from
${\bf{H}}_m$ by (\ref{equ:Jul3Add2}) is the $m\times m$ leading principal submatrix of ${\bf{R}}$ \cite{zhf3}. Then
we have
\begin{equation}\label{equ:15}
{\bf{R}}_{m}=\left[ {\begin{array}{*{20}c}
   {{\bf{R}}_{m - 1}} & {{\bf{v}}_{m - 1} }  \\
   { {\bf{v}}_{m - 1}^H } & {\beta _m }  \\
\end{array}} \right].
\end{equation}

Now let us substitute (\ref{equ:15}) and (\ref{Apd1AugEle1}) into
\begin{equation}{\label{ToBeSubstedforGama}}
{{\bf{F }}_{m}^{ - H} }  {{\bf{F }}_{m} ^{ - 1}}   = {\bf{R}}_{m},
\end{equation}
which is 
 deduced from (\ref{equ:8}) and (\ref{equ:3}).
Then we obtain
\begin{equation}{\label{Apd1AugEle2}}
\left[ \setlength{\arraycolsep}{0.7mm}
\renewcommand{\arraystretch}{1.5}
{\begin{array}{*{20}c}
    \times  & { - \frac{{{\bf{F}}_{m - 1}^{ - H} {\bf{F}}_{m - 1}^{ - 1} {\bf{u}}_{m - 1} }}{{\lambda _m }}}  \\
    \times  & {\frac{{{\bf{u}}_{m - 1}^H {\bf{F}}_{m - 1}^{ - H} {\bf{F}}_{m - 1}^{ - 1} {\bf{u}}_{m - 1}  + 1}}{{\lambda _m \lambda _m^* }}}  \\
\end{array}} \right] = \left[ {\begin{array}{*{20}c}
   {{\bf{R}}_{m - 1} } & {{\bf{v}}_{m - 1} }  \\
   {{\bf{v}}_{m - 1}^H } & {\beta _m }  \\
\end{array}} \right],
\end{equation}
where ``$\times$" denotes irrelevant entries.
From (\ref{Apd1AugEle2}), we deduce
\begin{subnumcases}{\label{derive4}}
- {{{\bf{F }}_{m-1}^{ - H} {\bf{F }}_{m-1}^{ -
1} {\bf{u}}_{m - 1} }}/{{\lambda _m }} = {\bf{v}}_{m - 1}, & \label{derive4a}\\
({{{\bf{u}}_{m - 1}^H {\bf{F }}_{m-1}^{ -
H} {\bf{F }}_{m-1}^{ - 1} {\bf{u}}_{m - 1} +
1}})/({{\lambda _m \lambda _m^* }})=\beta _m.  &  \label{derive4b}
\end{subnumcases}

From (\ref{derive4}), finally we can derive
\begin{subnumcases}{\label{equ20_19}}
\lambda _m  = 1/\sqrt {{{\beta _m  - {\bf{v}}_{m - 1}^H
{\bf{F }}_{m-1}{{\bf{F }}_{m-1}^H}
{\bf{v}}_{m - 1} }}}, &\label{equ:20}\\
{\bf{u}}_{m - 1}  =  - \lambda _m {\bf{F }}_{m-1} {{\bf{F
}}_{m-1}^H}{\bf{v}}_{m - 1}. &\label{equ:19}
\end{subnumcases}
We derive (\ref{equ:19}) from (\ref{derive4a}). 
Then
(\ref{equ:19}) is substituted
into (\ref{derive4b}) to derive
\begin{equation}\label{equ:18}
\lambda _m \lambda _m^*  = {\left({\beta _m  - {\bf{v}}_{m - 1}^H
{\bf{F }}_{m-1}  {{\bf{F }}_{m-1}^H} {\bf{v}}_{m - 1}
}\right)^{-1}},
\end{equation}
while
a $\lambda _m$ satisfying (\ref{equ:18})
can be computed by (\ref{equ:20}).

We can use (\ref{equ20_19}) and (\ref{equ:13}) to compute $ {\bf{F
}}_{m} $
 from ${\bf{F }}_{m-1}$
 iteratively till we get ${\bf{F }}_{M}$. The iterations start from $ {\bf{F
}}_{1} $ 
satisfying (\ref{ToBeSubstedforGama}), which
can be computed by
\begin{equation}\label{equ:22}
{\bf{F }}_{1}  = \sqrt {{{\bf{R}}_{1}^{ - 1} }}.
\end{equation}
Correspondingly instead of step P1,
we can propose 
step N1 to compute an initial
upper-triangular ${\bf{F }}$,
which includes the
following sub-steps.

\line(1,0){240} 

\qquad\qquad{\bf The Sub-steps of Step N1}

\line(1,0){240} 
\begin{enumerate}
\item[N1-a)] Assume 
the successive detection order
to be
 $t_M, t_{M - 1}, \cdots, t_1$. 
Correspondingly permute 
$\bf H$ to be ${\bf{H}}={\bf{H}}_M = [{\bf{h}}_{t_1},{\bf{h}}_{t_2},
\cdots,{\bf{h}}_{t_M}]$, and permute 
${\bf{a}}$ to be ${\bf{a}}={\bf{a}}_M= [a_{t_1},a_{t_2},
\cdots,a_{t_M}]^T$. 
\item[N1-b)] Utilize the permuted $\bf H$ to compute ${\bf{R}}_{M}$, where 
we can obtain all
 ${\bf{R}}_{m - 1}$s, ${\bf{v}}_{m - 1}$s and $\beta _{m}$s \cite{zhf3} (for $m=M,M-1,\cdots,2$),
 as shown in (\ref{equ:15}).
\item[N1-c)] Compute ${\bf{F}}_{1}$
 by (\ref{equ:22}). Then use (\ref{equ20_19}) and (\ref{equ:13})
 to compute ${\bf{F}}_{m}$
 from ${\bf{F}}_{m - 1}$
 iteratively for $m=2,3,\cdots,M$, to obtain the initial $ {\bf{F }}={\bf{F}}_{M}$.
\end{enumerate}
\line(1,0){252} 

 The obtained upper triangular ${\bf{F
  }}_{M}$  is equivalent to a Cholesky factor \cite{zhf_VTC08_6} of
   ${\bf{P}}_{M}={\bf{R}}_{M}^{-1}$, since 
   ${\bf{F}}_{M}$
   and
   ${\bf{P}}_{M}$ can be permuted to the lower
triangular ${\bf{F
 }}_{M}$ and the corresponding ${\bf{P}}_{M}$, which still satisfy (\ref{equ:8}). Notice that the ${\bf{F
}}_{M}$ with columns exchanged still satisfies (\ref{equ:8}), while
if two rows in ${\bf{F }}_{M}$ are exchanged, the corresponding two
rows and columns in ${\bf{P}}_{M}$ need to be exchanged.

Now from (\ref{equ:15}), (\ref{equ:8}) and (\ref{equ:3}), it can
be seen
that (\ref{equ:9})
(proposed in \cite{zhf2})
 and (\ref{equ:13}) actually reveal the relation between the $m^{th}$ and
the $(m-1)^{th}$ order inverse Cholesky factor of the matrix
${\bf{R}}$. This relation is also utilized
to implement adaptive filters
in \cite{reviewer_VTC08_e,reviewer_VTC08_d},
where the $m^{th}$ order
inverse Cholesky factor is obtained from the $m^{th}$ order Cholesky
factor \cite[equation (12)]{reviewer_VTC08_e}, \cite[equation
(16)]{reviewer_VTC08_d}.
Thus the algorithms in
\cite{reviewer_VTC08_e,reviewer_VTC08_d} are still similar to the
conventional matrix inversion algorithm \cite{FixedPointChskInv}
using Cholesky factorization, where the inverse Cholesky factor is
computed from the Cholesky factor by the back-substitution (for
triangular matrix inversion), an inherent serial process unsuitable
for the parallel implementation \cite{BackSubstiteNoParellel}.
Contrarily, 
the proposed algorithm computes
 the 
 inverse Cholesky factor of
${\bf{R}}_{m} $ 
 from ${\bf{R}}_{m}$ directly, as shown in (\ref{equ20_19}) and
(\ref{equ:13}). Then it can
 avoid the conventional back substitution  of the
Cholesky factor.

In a word, although the relation between the $m^{th}$ and the
$(m-1)^{th}$ order inverse Cholesky factor (i.e. (\ref{equ:9}) and
(\ref{equ:13})) has been mentioned
\cite{zhf2,reviewer_VTC08_e,reviewer_VTC08_d}, our contributions in
this letter include substituting this relation into
(\ref{ToBeSubstedforGama})
to find (\ref{equ:18}) and
(\ref{equ20_19}). 
Specifically,
to compute the $m^{th}$ order inverse Cholesky
factor, the conventional matrix inversion algorithm using Cholesky
factorization \cite{FixedPointChskInv} usually requires $2m$ divisions (i.e.
$m$ divisions for Cholesky factorization and the other $m$ divisions
for the back-substitution), while the 
proposed algorithm
only requires  $m$ divisions to compute (\ref{equ:22})
and (\ref{equ:20}).

\section{The Proposed Square-root V-BLAST Algorithm}
Now ${\bf{R}}_{M}$ has been computed in sub-step N1-b. Thus as the
recursive V-BLAST algorithm in \cite{zhf4}, we can also cancel the
interference of the detected signal $a_m$ in
\begin{equation} \label{equ:23}
{\bf{z}}_{m}={{\bf{H}}_{m} ^H}{\bf{x}}^{(m)}
\end{equation}
by
\begin{equation} \label{equ:25}
{\bf{z}}_{m - 1}  = {\bf{z}}_{m}^{[-1]}  - a_m  \cdot {\bf v}_{m-1},
\end{equation}
where $ {\bf{z}}_{m}^{[-1]}$ is the permuted ${\bf{z}}_{m}$ with the
last entry removed, and ${\bf v}_{m-1}$ is
in the permuted ${\bf{R}}_{m}$ \cite{zhf4,zhf6}, as shown in
(\ref{equ:15}). Then to
avoid computing ${{\bf{H}}_{m}^H }{\bf{x}}^{(m)}$ in (\ref{equ:10}), 
we form the estimate of $a_m$ by
 \begin{equation} \label{equ:24}
\hat{a}_m  = \lambda _m  \cdot \left[ {\begin{array}{*{20}c}
   {\left( {{\bf{u}}_{m - 1} } \right)^H } & {\left( {\lambda _m } \right)^* }  \\
\end{array}} \right] \cdot {\bf{z}}_{m}.
\end{equation}

It is required to compute the initial ${\bf{z}}_{M}$. So step N1 should include the following sub-step N1-d. 
\begin{enumerate}
\item[N1-d)] Compute $
{\bf{z}}_{M}={{\bf{H}}_{M}^H} {\bf{x}}^{(M)}={{\bf{H}}_{M}^H}{\bf{x}}$.
\end{enumerate}

The proposed square-root V-BLAST algorithm is summarized
as follows.

\line(1,0){240}

\qquad {\bf The Proposed Square-root V-BLAST Algorithm}

\line(1,0){240}

{\sl Initialization}:
\begin{enumerate}
\item[N1)] Set $m=M$. Compute ${\bf{R}}_{M} $,  ${\bf{z}}_{M}$  
and the initial upper triangular $ {\bf{F }}  = {\bf{F
}}_{M} $. This step includes the above-described sub-steps N1-a,
N1-b, N1-c and N1-d.
\end{enumerate}

{\sl{Iterative Detection}}:
\begin{enumerate}
\item[N2)] Find the minimum length row in ${\bf{F }}_{m} $
 and permute it to be the last $m^{th}$ row. Correspondingly permute $
{\bf{a}}_{m} $, $ {\bf{z}}_{m} $,
 and rows and columns in ${\bf{R}}_{m}$ \cite{zhf6}.
\item[N3)] Block upper-triangularize ${\bf{F }}_{m}$ by
 (\ref{equ:9}).
\item[N4)] Form the least-mean-square estimate $ \hat{a}_m $ by (\ref{equ:24}).
\item[N5)] Obtain $a_m$ from $\hat{a}_m$ via slicing.
\item[N6)] Cancel the effect of $a_m$ in ${\bf{z}}_{m}$
 by (\ref{equ:25}), to obtain the reduced-order problem (\ref{EqAug1RedOrdPrb}) with the corresponding
 ${\bf{z}}_{m - 1}$, ${\bf{a}}_{m - 1}$, ${\bf{R}}_{m - 1}$ and ${\bf{F }}_{m-1}$.
\item[N7)] If $m>1$, let $m=m-1$ and go back to step N2. 
\end{enumerate}
\line(1,0){252}

Since ${\bf{F}}_{M}$ obtained in step N1 is upper triangular,
step N3 requires less computational load than the corresponding step
P3 (described in Section \Rmnum{3}),
which is 
analyzed as follows. ¡¡

Suppose that the minimum length row of $ {\bf{F }}_{M} $ found in 
step N2 is the $i^{th}$ row, which
must be
\begin{displaymath}
\left[ {\begin{array}{*{20}c}
   0 &  \cdots  & 0 & {f_{ii} }  & \cdots  & {f_{iM}}  \\
\end{array}} \right]
\end{displaymath}
with the first
$i-1$ entries to be zeros. 
 Thus in step N3 the transformation $
{\bf{\Sigma }} $ can be performed by only $(M-i)$ Givens rotations
\cite{zhf_VTC08_6}, i.e.,
\begin{equation}\label{equ:26}
{\bm \Sigma}^g_M = {\bf{\Omega }}_{i,i + 1}^i {\bf{\Omega }}_{i +
1,i + 2}^i \cdots {\bf{\Omega }}_{M - 1,M}^i=\prod\limits_{j = i}^{M
- 1} {{\bf{\Omega }}_{j,j + 1}^i },
\end{equation}
where the Givens rotation ${\bf{\Omega }}_{k,n}^i$
 rotates the $k^{th}$ and $n^{th}$ entries 
 in each row of ${\bf{F }}_{M}$,
 and zeroes the $k^{th}$  entry in the $i^{th}$ row.

In step N2, we can delete the $i^{th}$ row in ${\bf{F }}_{M}$
firstly to get ${\bf{\bar F}}_{M}$, and then
add the deleted $i^{th}$ row to ${\bf{\bar F}}_{M}$ as the last row
to obtain the permuted ${\bf{F }}_{M}$.
Now
it is easy to verify that the ${\bf{F }}_{M-1}$ obtained from
${\bf{F }}_{M} {\bm \Sigma}^g_M$
by (\ref{equ:9}) is still upper triangular. 
For the subsequent $m=M-1,M-2,\cdots,2$, we also obtain
${\bf{F }}_{m-1}$ from ${\bf{F }}_{m} {\bm \Sigma}^g_m$
by (\ref{equ:9}), where ${\bm
\Sigma}^g_m$ is defined by (\ref{equ:26}) with $M=m$.
Correspondingly
we can deduce
that ${\bf{F }}_{m-1}$ 
is also triangular. Thus ${\bf{F }}_{m}$ is always triangular, for $m=M,M-1,\cdots,1$.

To sum up, our contributions in this letter include steps N1, N3, N4
and N6 that improve steps P1, P3, P4 and P6 (of the previous
square-root V-BLAST algorithm \cite{vtc08_zhfVTC4}), respectively.
  Steps N4 and N6 come from the extension of the 
  improvement 
  in \cite{zhf4} (for the recursive V-BLAST algorithm) to the square-root V-BLAST
algorithm. 
However, it is infeasible to extend the improvement in \cite{zhf4}
to the existing square-root V-BLAST algorithms in
\cite{zhf2,vtc08_zhfVTC4}, since they do not provide ${\bf{R}}_{M}$
that is required to get ${\bf v}_{m-1}$ for (\ref{equ:25}).

\section{Complexity Evaluation}
In this section, 
($j$, $k$) denotes
the computational complexity of $j$ complex
multiplications and $k$ complex additions, which is simplified 
 to ($j$) if $j=k$. Similarly, $\left\langle {{\chi}_1,{\chi}_2,{\chi}_3} \right\rangle$ denotes that the
speedups in the number of multiplications,  additions and
floating-point operations (flops)
are ${\chi}_1$, ${\chi}_2$ and ${\chi}_3$, respectively, which is
simplified
to $\left\langle {{\chi}_1} \right\rangle$ if
${\chi}_1={\chi}_2={\chi}_3$. Table \Rmnum{1} compares the expected
complexity of the proposed square-root V-BLAST algorithm
  and that of the previous
  one in
  \cite{vtc08_zhfVTC4}.
The detailed complexity derivation is as follows.

In sub-step N1-c, the dominant computations come from
(\ref{equ20_19}). It needs a complexity of
$\left(\frac{(m-1)m}{2}\right)$ to compute
${\bf{y }}_{m - 1}  = {\bf{F }}_{m-1}^H {\bf{v}}_{m - 1}$
firstly, where ${\bf{F }}_{m-1}$ is triangular. Then to obtain the
$m^{th}$ column of ${\bf{F }}$, we
compute
(\ref{equ20_19}) by 
\begin{subnumcases}{\label{yToLamda_u}}
  \lambda _m = \sqrt {1/\left( {\beta _m -
 {\bf{y }}_{m - 1}^H {\bf{y }}_{m - 1} } \right)}, & \label{yToLamda}\\
 {\bf{u}}_{m - 1}  =  - \lambda _m
{\bf{F }}_{m-1} {\bf{y }}_{m - 1}. &  \label{yTo_u}
\end{subnumcases}
In (\ref{yToLamda_u}), the complexity to compute ${\bf{F }}_{m-1}
{\bf{y }}_{m - 1}$ is $\left(\frac{(m-1)m}{2}\right)$, and that to
compute the other parts is $\left(O(m)\right)$.
So sub-step N1-c totally requires a complexity of
$\left(\sum\limits_{m=2}^M {\frac{(m-1)m}{2}\times2}+O(m)\right) =
\left(\frac{{M^3 }}{3}+O(M^2)\right)$ to compute (\ref{equ20_19})
for $M-1$ iterations, while sub-step N1-b requires a complexity of
($\frac{M^{2}N}{2}$) \cite{zhf3} to compute the Hermitian
${\bf{R}}_{M}$. 
As a comparison,
in each of the $N(>M-1)$ iterations, step P1
computes ${{\bf{\underline h}}_i^{H} {\bf{\underline P}}_{i -
1}^{1/2}}$ to form the $(M+1)\times (M+1)$ pre-array ${\bf{\Pi}}_i$,
and then block lower-triangularizes ${\bf{\Pi}}_i$ by the
$(M+1)\times (M+1)$ Householder transformation \cite{vtc08_zhfVTC4}.
Thus it can be seen that step P1 requires much more complexity than
the proposed step N1.



In steps N3 and P3, we can apply the efficient complex Givens rotation
\cite{Complex_Givens_Jul14b}
%
${\bf{\Phi }} =\frac{1}{q}\left[ {\begin{array}{*{20}c}
   c & s  \\
   { - s^* } & c  \\
\end{array}} \right]$ to rotate $\left[
{\begin{array}{*{20}c}
   d & e  \\
\end{array}} \right]$ into  $\left[ {\begin{array}{*{20}c}
   0 & {\left( {e/\left| e \right|} \right)q}  \\
\end{array}} \right]$,  where
$c = \left| e \right|$ and $q = \sqrt {\left| e \right|^2  + \left|
d \right|^2 } $ are real, and $s = \left( {e/\left| e \right|}
\right)d^*$ is complex. The efficient Givens rotation equivalently requires \cite{zhf3} $3$
complex multiplications and $1$ complex additions to rotate a row.
Correspondingly the complexity of step P3 is $(M^{3},\frac{1}{3}M^{3})$.
Moreover,  step P3 can also adopt a Householder reflection, and then requires a complexity of
$(\frac{2}{3}M^{3})$ \cite{vtc08_zhfVTC4}.
On the other hand, the Givens rotation ${{\bf{\Omega }}_{j,j + 1}^i }$ in (\ref{equ:26}) only rotates
non-zero entries in the first $j+1$ rows of the upper-triangular
${\bf{F}}_{M}$. Then (\ref{equ:26}) requires a complexity of
$\left(\sum\limits_{j = i}^{m-1}{3(j+1)}\approx
\frac{{3(m^2-i^2)}}{2},\frac{{(m^2-i^2)}}{2}\right)$. When the
detection order assumed in sub-step N1-a is statistically
independent of the optimal detection order, the probabilities for
$i=1, 2, \cdots, m$ are equal. Correspondingly the expected (or
average)
complexity of step N3 
is $\left(\sum\limits_{m = 1}^M {{\frac{1}{m}\sum\limits_{i = 1}^m
{\frac{{3(m^2-i^2)}}{2}}}} \approx \frac{{M^3 }}{3},\frac{{M^3
}}{9}\right)$. Moreover, when the probability for $i=1$ is 100\%,
step N3 needs the worst-case complexity, which is
$\left(\sum\limits_{m = 1}^M { { {\frac{{3(m^2-1^2)}}{2}}}} \approx
\frac{{M^3 }}{2},\frac{{M^3 }}{6}\right)$. Correspondingly
we can
deduce that the worst-case complexity of the proposed
V-BLAST algorithm is
$(\frac{2}{3}M^{3}+\frac{M^{2}N}{2},\frac{M^{3}}{3}+\frac{7}{2}M^{2}N)-(
\frac{{M^3 }}{3},\frac{{M^3 }}{9})+(\frac{{M^3 }}{2},\frac{{M^3
}}{6})=(\frac{5}{6}M^{3}+\frac{1}{2}M^{2}N,\frac{1}{2}M^{3}+\frac{1}{2}M^{2}N)$.
The ratio between the worst-case and expected flops of the proposed
square-root algorithm is only $1.125$, while recently there is a
trend to study the expected, rather than worst-case, complexity of
various
algorithms \cite{Hassibi_Sphere_1}. 
Thus only the expected complexity is considered in Table \Rmnum{1}
and in what follows.

In MIMO OFDM systems, the 
complexity of step N3 can be further reduced, and can even be zero.
In sub-step N1-a, we assume the detection order
to be 
the optimal order of the adjacent subcarrier, which is quite similar
or even identical to the actual optimal detection order
\cite{zhf_VTC08_8}. 
Correspondingly the required Givens rotations are less or even zero.
So the expected complexity of step N3 ranges from
 $(\frac{1}{3}M^{3},\frac{1}{9}M^{3})$ to 
 zero,
 while the exact mean value depends on the statistical correlation between the assumed detection order
 and the actual optimal detection order.

The complexities
of the ZF-OSIC V-BLAST algorithm in \cite{vtc08_reviewer_a} and the
MMSE-OSIC V-BLAST algorithms in
\cite{zhf2,zhf3,reviewer_VTC08_b,zhf6}
are
$(\frac{1}{2}M^{3}+2M^{2}N)$,
$(\frac{2}{3}M^{3}+4M^{2}N+MN^{2})$\cite{vtc08_zhfVTC4},
$(\frac{2}{3}M^{3}+3M^{2}N,\frac{1}{2}M^{3}+\frac{5}{2}M^{2}N)$,
$(\frac{2}{3}M^{3}+\frac{5}{2}M^{2}N)$ and
$(\frac{2}{3}M^{3}+\frac{1}{2}M^{2}N)$, respectively. 
Let $M=N$. Also assume the
transformation ${\bf{\Sigma}}$ in \cite{vtc08_zhfVTC4} to be a
sequence of efficient Givens rotations \cite{Complex_Givens_Jul14b}
that are hardware-friendly \cite{zhf2}.
Then the expected speedups of the proposed square-root algorithm
over
the
previous
one \cite{vtc08_zhfVTC4} range from $\left\langle
{\frac{9}{2}/\frac{7}{6}=3.86,\frac{23}{6}/\frac{17}{18}=4.06,3.9}
\right\rangle$ to $\left\langle
{\frac{9}{2}/\frac{5}{6}=5.4,\frac{23}{6}/\frac{5}{6}=4.6,5.2}
\right\rangle$, while the expected speedups of the proposed
algorithm over
the fastest known recursive algorithm \cite{zhf6} range from
$\left\langle
{\frac{7}{6}/\frac{7}{6}=1,\frac{7}{6}/\frac{17}{18}=1.24,1.05}
\right\rangle$ to $\left\langle {\frac{7}{6}/\frac{5}{6}=1.4}
\right\rangle$.

For more fair comparison,
 we 
 modify
the fastest known recursive
algorithm \cite{zhf6} 
to further reduce the complexity. We spend extra memories to store
each intermediate ${\bf{P }}_{m}$ ($m=1,2,\cdots,M-1$) computed in
the {\sl initialization} phase, which may be equal to the ${\bf{P
}}_{m}$ required in the {\sl recursion} phase \cite{zhf6}. Assume
the successive detection order and permute $\bf H$ accordingly, as
in sub-step N1-a. When the assumed
order is identical to the actual optimal detection order, 
each ${\bf{P }}_{m}$ 
required in the {\sl
recursion} phase is equal to the stored ${\bf{P }}_{m}$. 
Thus we can achieve the maximum complexity savings, i.e. the
complexity of $\left(\frac{1}{6}M^{3}+O(M^{2})\right)$
\cite[equations (23) and (24)]{zhf6}
to deflate ${\bf{P }}_{m}$s.
On the other hand, when the assumed order is statistically
independent of the actual optimal detection order, there is an equal
probability for the $m$ undetected antennas to be any of 
the ${C_m^M}$ possible antenna combinations. Correspondingly
$1/{C_m^M}=\frac{{(M - m)!m!}}{{M!}}$ is the probability for the
stored ${\bf{P }}_{m}$ to be equal to the ${\bf{P }}_{m}$ required
in the {\sl recursion} phase. Thus
we can obtain the minimum expected complexity savings, i.e.  
\cite[equations (23) and (24)]{zhf6},
\begin{equation}\label{saveOFrecurAlg}
\left( \sum\limits_{m = 2}^M
{{\frac{1}{C_m^M}}{\frac{(m-1)(m+2)}{2}}}, \sum\limits_{m = 2}^M
{{\frac{1}{C_m^M}}{\frac{(m-1)m}{2}}}\right).
\end{equation}
The ratio of the minimum expected complexity savings
to the maximum complexity savings 
is 22\%
when $M=4$, and is only 
1.2\% 
when $M=16$. 
It can be seen that
the minimum expected complexity savings are negligible when $M$ is large. 
The minimum complexity of the recursive V-BLAST algorithm \cite{zhf6} with the above-described modification,
which is $(\frac{2}{3}M^{3}+\frac{1}{2}M^{2}N-\frac{1}{6}M^{3})=(\frac{1}{2}M^{3}+\frac{1}{2}M^{2}N)$, is still more than
that of the proposed square-root V-BLAST algorithm.
When $M=N$, the ratio of the former to
the latter is ${1/\frac{5}{6}=1.2}$.

Assume $M=N$. For different number of transmit/receive antennas, we
carried out some numerical experiments to count the average flops of
  the OSIC V-BLAST algorithms in
\cite{zhf2,
zhf3,reviewer_VTC08_b,zhf6,vtc08_reviewer_a,vtc08_zhfVTC4}, the
proposed square-root V-BLAST algorithm, and the recursive V-BLAST algorithm
\cite{zhf6} with the above-described modification. The results are shown in Fig. 1.
It can be seen that they are consistent with the theoretical flops
calculation.
%

\section{Conclusion}
We propose a  fast algorithm for inverse Cholesky factorization, to compute a triangular square-root of the estimation error covariance matrix for V-BLAST. Then it is employed to propose
an improved square-root algorithm for V-BLAST, which speedups several steps in the previous
one \cite{vtc08_zhfVTC4}, and 
 can offer further
computational savings
 in MIMO OFDM systems.
Compared to the
conventional inverse Cholesky factorization, the proposed one 
avoids
the back substitution (of the
Cholesky factor), an inherent serial process unsuitable
for the parallel implementation \cite{BackSubstiteNoParellel}, and then 
requires only half divisions. 
The proposed 
V-BLAST algorithm is faster than the
existing efficient V-BLAST algorithms in
\cite{zhf2,vtc08_zhfVTC4,vtc08_reviewer_a,zhf3,zhf5,zhf4,reviewer_VTC08_b,zhf6}.
Assume $M$ transmitters and 
the equal number of
receivers.
In MIMO OFDM systems, the expected speedups (in the number of flops)
of the proposed square-root V-BLAST algorithm over the
previous 
 one
 \cite{vtc08_zhfVTC4} and the fastest known recursive V-BLAST algorithm \cite{zhf6} are
$3.9\sim5.2$ and $1.05\sim1.4$, respectively.
The recursive
algorithm \cite{zhf6} can be modified to further reduce the
complexity at the price of extra memory consumption, while the
minimum expected complexity savings are negligible when $M$ is
large. The speedups of the proposed square-root algorithm over the
fastest known recursive algorithm \cite{zhf6} with the
above-mentioned modification are $1.2$, when both algorithms are
assumed to achieve the maximum complexity savings.
Furthermore, as shown in \cite{MyVTC2010spring}, the proposed square-root algorithm 
can also be applied in
the extended V-BLAST with selective per-antenna rate control (S-PARC), to reduce the complexity even by a factor of $M$.
\appendices

%

\ifCLASSOPTIONcaptionsoff
  \newpage
\fi

\newpage

\begin{table}[!t]
\renewcommand{\arraystretch}{1.3}
\caption{Complexity Comparison between the Proposed Square-Root
V-BLAST Algorithm and the Previous Square-Root V-BLAST Algorithm in
\cite{vtc08_zhfVTC4}} \label{table_example} \centering
\begin{tabular}{c|c|c|}
\bfseries  Step   & \bfseries The Algorithm in \cite{vtc08_zhfVTC4}  & \bfseries The Proposed Algorithm \\
\hline
\bfseries  \bfseries 1-b & ($3M^{2}N$) \cite{vtc08_zhfVTC4} for step P1 & ($\frac{M^{2}N}{2}$)\cite{zhf3}  \\
\bfseries  \bfseries 1-c &  & ($\frac{M^{3}}{3}$)  \\
\hline
\bfseries  \bfseries 3 & ($M^{3}, \frac{M^{3}}{3}$) or ($\frac{2}{3}M^{3}$) \cite{vtc08_zhfVTC4} & From ($\frac{M^{3}}{3}, \frac{M^{3}}{9}$) to ($0$)  \\
\hline
\bfseries  \bfseries 4 & ($\frac{M^{2}N}{2}$) \cite{vtc08_zhfVTC4} & $\left(0( M^{3})\right)$  \\
\hline
\bfseries  \bfseries Sum & ($M^{3}+\frac{7}{2}M^{2}N,$   &From $(\frac{2}{3}M^{3}+\frac{M^{2}N}{2},$   \\
\bfseries  \bfseries       & $\frac{M^{3}}{3}+\frac{7}{2}M^{2}N$) & $\frac{4}{9}M^{3}+\frac{M^{2}N}{2})$ \\
\bfseries  \bfseries       & or ($\frac{2}{3}M^{3}+\frac{7}{2}M^{2}N$) & to ($\frac{M^{3}}{3}+\frac{M^{2}N}{2}$)\\
\hline
\end{tabular}
\end{table}

\begin{figure}[!t]
\centering
\includegraphics[width=6 in]{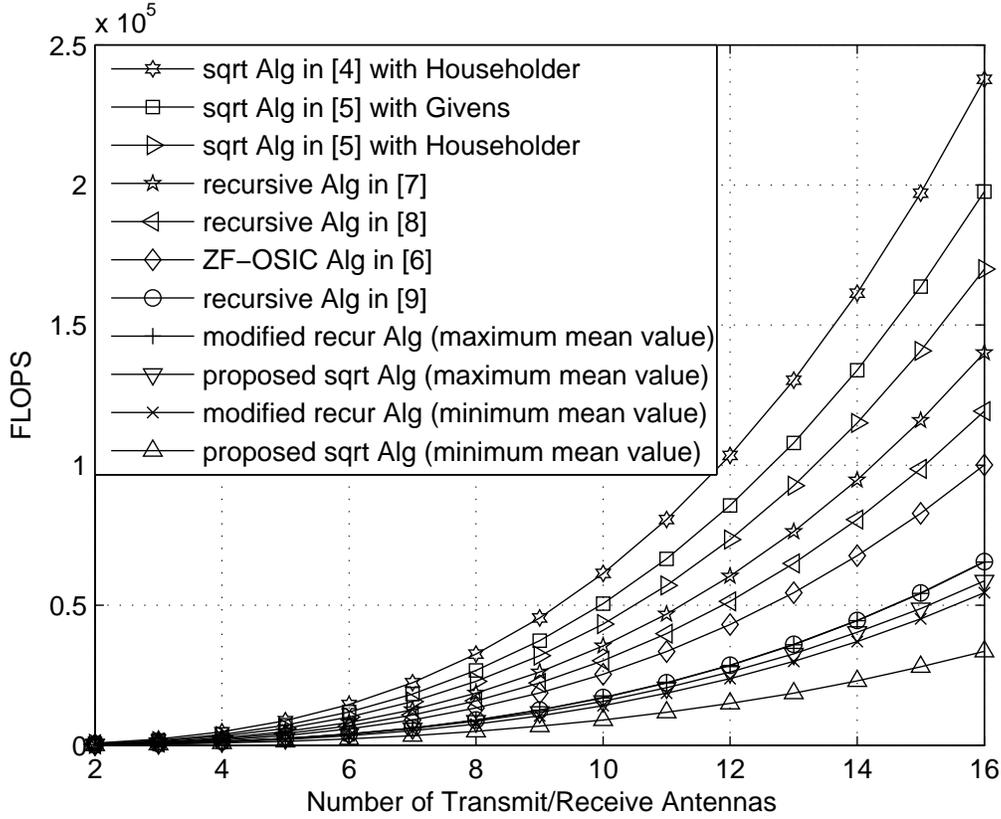}
\caption{Comparison of computational complexities among the
MMSE-OSIC algorithms in
\cite{zhf2,vtc08_zhfVTC4,zhf3,reviewer_VTC08_b,zhf6} and this
letter, and the ZF-OSIC algorithm
 in \cite{vtc08_reviewer_a}. 
 ``sqrt" and ``Alg"
means square-root and algorithm, respectively. ``$\cdots$ with
Householder" and ``$\cdots$ with Givens" adopt a Householder
reflection and a sequence of Givens rotations, respectively.
Moreover, ``modified recur Alg" is the recursive algorithm
\cite{zhf6} with the modification described in this letter.}
\label{fig_sim}
\end{figure}


\begin{thebibliography}{1}

\bibitem{Jun28_MIMO}  G. J. Foschini and M. J. Gans,
``On limits of wireless communications in a fading environment when
using multiple antennas," \emph{Wireless Personal Commun.}, pp.
311-335, Mar. 1998.

\bibitem{BLASTfeb25}  G. J. Foschini,
``Layered space-time architecture for wireless communication
in a fading environment using multi-element antennas," \emph{Bell Labs.
Tech. J.,}, vol. 1, no. 2, pp. 41¨C59, 1996.

\bibitem{zhf1} P. W. Wolniansky, G. J. Foschini, G. D. Golden and R. A. Valenzuela,
``V-BLAST: an architecture for realizing very high data rates over
the rich-scattering wireless channel", \emph{Proc. Int. Symp.
Signals, Syst., Electron. (ISSSE¡¯98)}, pp. 295-300, Sept. 1998.

\bibitem{zhf2} B. Hassibi, ``An efficient square-root algorithm for BLAST",
\emph{Proc. IEEE Int. Conf. Acoustics, Speech, and Signal
Processing, (ICASSP '00)}, pp. 737-740, June 2000.

 \bibitem{vtc08_zhfVTC4}
H. Zhu, Z. Lei and F. P. S. Chin, ``An improved square-root
algorithm for BLAST",  \emph{IEEE Signal Processing Letters}, vol.
11, no. 9, pp. 772-775, Sept. 2004.

\bibitem{vtc08_reviewer_a}
 W. Zha and S. D. Blostein, ``Modified decorrelating decision-feedback detection
 of BLAST space-time system", \emph{Proc. IEEE ICC}, vol. 4,  pp. 59-63, 2002.


\bibitem{zhf3} J. Benesty, Y. Huang and J. Chen, ``A fast recursive algorithm for
optimum sequential signal detection in a BLAST system", \emph{IEEE
Trans. on Signal Processing}, pp. 1722-1730, July 2003.

\bibitem{reviewer_VTC08_b}
Z. Luo, S. Liu, M. Zhao and Y. Liu, ``A Novel Fast Recursive
MMSE-SIC Detection Algorithm for V-BLAST Systems", \emph{IEEE
Transactions on Wireless Communications}, vol. 6,  Issue 6, pp. 2022 - 2025, June
2007.

\bibitem{zhf6} Y. Shang and X. G. Xia, ``On Fast Recursive Algorithms For V-BLAST With Optimal Ordered SIC Detection",
 \emph{IEEE Transactions on Wireless Communications}, vol. 8, pp. 2860-2865, June 2009.

\bibitem{zhf5}  L. Szczeci¡änski, and D. Massicotte, ``Low complexity adaptation of MIMO MMSE receivers,
Implementation aspects", \emph{Proc. IEEE Global Commun. Conf.
(Globecom¡¯05)}, St. Louis, MO, USA, Nov. 28 - Dec. 2, 2005.

\bibitem{zhf4} H. Zhu, Z. Lei, and F. P. S. Chin, ``An improved recursive algorithm for
BLAST", \emph{Signal Process.}, vol. 87, no. 6, pp. 1408-1411, Jun.
2007.






\bibitem{OFDM2_zhf_VTC08_8} N. Boubaker, K.B.
Letaief and R.D. Murch, ``A low complexity multicarrier BLAST
architecture for realizing high data rates over dispersive fading
channels,"  \emph{IEEE Vehicular Technology Conference (VTC), 2001
Spring}, May 2001.


\bibitem{zhf_VTC08_8}
W. Yan, S. Sun and Z. Lei, ``A low complexity VBLAST OFDM detection
algorithm for wireless LAN systems",
 \emph{IEEE Communications Letters}, vol. 8, no. 6, pp. 374-376, June 2004.

 \bibitem{zhf_VTC08_6}
 G. H. Golub and C. F. Van Loan, \emph{Matrix Computations}, Johns Hopkins University Press,
 Baltimore, MD, 3rd edition, 1996.



 \bibitem{reviewer_VTC08_e}
A. A. Rontogiannis and S. Theodoridis, ``New fast QR decomposition
least squares adaptive algorithms", \emph{IEEE Trans. on Signal
Processing}, vol. 46, no. 8, pp. 2113-2121, Aug. 1998.

 \bibitem{reviewer_VTC08_d}
 A. A. Rontogiannis, V. Kekatos and K. Berberidis, ``A square-root
adaptive V-BLAST algorithm for fast time-varying MIMO channels",
\emph{IEEE Signal Processing Letters}, vol. 13, no. 5, pp. 265-268,
May 2006.







%
%
%
%
%
%




\bibitem{FixedPointChskInv}
A. Burian, J. Takala and M. Ylinen, ``A fixed-point implementation
of matrix inversion using Cholesky decomposition",
\emph{Micro-NanoMechatronics and Human Science, 2003 IEEE
International Symposium on}, 27-30 Dec. 2003, vol. 3, pp. 1431-1434.


\bibitem{BackSubstiteNoParellel}
E. J. Baranoski, ``Triangular factorization of inverse data
covariance matrices",  \emph{International Conference on Acoustics, Speech, and Signal
Processing, 1991 (ICASSP-91)},
14-17 Apr 1991, pp. 2245 - 2247, vol.3.


 \bibitem{Complex_Givens_Jul14b}
D. Bindel, J. Demmel, W. Kahan and O. Marques, ``On Computing Givens
rotations reliably and efficiently", \emph{ACM Transactions on
Mathematical Software (TOMS) archive}, vol. 28 , Issue 2, June 2002.
Available online at: www.netlib.org/lapack/lawns/downloads/.



 \bibitem{Hassibi_Sphere_1}
B. Hassibi and H. Vikalo, ``On the sphere decoding algorithm: Part I, the expected complexity", \emph{IEEE Transactions on Signal Processing}, vol 53, no 8, pages 2806-2818, Aug. 2005.



\bibitem{MyVTC2010spring}
 H. Zhu, W. Chen and B. Li, ``Efficient Square-root Algorithms for the Extended
V-BLAST with Selective Per-Antenna Rate Control", \emph{IEEE Vehicular Technology Conference (VTC), 2010
Spring}, May 2010.
\end{thebibliography}
\end{document}